\documentclass[traditabstract]{aa}

\usepackage{txfonts}

\usepackage{graphicx}

\usepackage{natbib}
\bibpunct{(}{)}{;}{a}{}{,} 

\def\solar {\ifmmode_{\mathord\odot} \else $_{\mathord\odot}$\fi}
\def\Msol {\ifmmode {\,{\it M}\solar} \else $\,M$\solar\fi}     
\def\Rsol {\ifmmode {\,{\it R}\solar} \else $\,R$\solar\fi}     
\def\Lsol {\ifmmode {\,{\it L}\solar} \else $\,L$\solar\fi}     
\newcommand{\Mjup}{M$_{Jup}$}

\begin{document}

\title{The HARPS search for southern extra-solar planets
       \thanks{Based on observations made with the HARPS instrument on the
         ESO 3.6-m telescope at La Silla Observatory under program ID
         072.C-0488}
       }
\subtitle{XXVIII. Two giant planets around M0 dwarfs }

\author{
T. Forveille \inst{1}
\and X. Bonfils \inst{1,2}
\and G. Lo Curto \inst{3}
\and X. Delfosse \inst{1}
\and S. Udry \inst{2}
\and F. Bouchy \inst{4,5}
\and C. Lovis \inst{2}
\and M. Mayor \inst{2}
\and C. Moutou \inst{6}
\and D. Naef\inst{3,2}
\and F. Pepe \inst{2}
\and C. Perrier \inst{1}
\and D. Queloz \inst{2}
\and N. Santos \inst{7,8}
}

\offprints{T. Forveille, \email{Thierry.Forveille@ujf-grenoble.fr}}

\institute{Laboratoire d'Astrophysique de Grenoble,
               Observatoire de Grenoble,
               Universit\'e Joseph Fourier,
               CNRS, UMR 5571,
               BP 53, 
               F-38041, Grenoble Cedex 9,
               France
\and
               Observatoire de Gen\`eve, 
               Universit\'e de Gen\`eve, 
               51 ch. des Maillettes, 
               1290 Sauverny,
               Switzerland
\and
               European Southern Observatory, 
               Alonso de Cordova 3107, 
               Vitacura, Santiago, Chile
\and  
	       Institut d'Astrophysique de Paris, 
               CNRS, Universit\'e Pierre et Marie Curie, 
               98bis Bd Arago, F-75014 Paris, France
\and
               Observatoire de Haute-Provence, 
               CNRS/OAMP, 
               F-04870 St Michel l'Observatoire, France
\and
               Laboratoire d'Astrophysique de Marseille, 
               38 rue Fr\'ederic Joliot-Curie, 
               F-13388 Marseille Cedex 13, France
\and
               Centro de Astrof{\'\i}sica, Universidade do Porto, 
               Rua das Estrelas, 
               4150-762 Porto, Portugal
\and
               Departamento de Física e Astronomia, 
               Faculdade de Ciências, Universidade do Porto,
               Rua das Estrelas, 
               4150-762 Porto, Portugal
}

\abstract{
Fewer giants planets are found around M dwarfs than around more 
massive stars, and this dependence of planetary characteristics 
on the mass of the central star is an important observational 
diagnostic of planetary formation theories. In part to improve
on those statistics, we are monitoring the radial velocities of 
nearby M dwarfs with the {\it HARPS} spectrograph on the ESO 
3.6~m telescope. We present here the detection of 
giant planets around two nearby M0 dwarfs: planets, with minimum
masses of respectively 5 Jupiter masses and 1 Saturn mass, orbit
around Gl~676A and HIP~12961.
The latter is, by over a factor of two, the most massive planet found
by radial velocity monitoring of an M dwarf, but its being found 
around an early M-dwarf is in approximate line with the upper 
envelope of the planetary vs stellar mass diagram. HIP~12961 
([Fe/H]=-0.07) is slightly more metal-rich than the average solar 
neighborhood ([Fe/H]=-0.17), and Gl~676A ([Fe/H=0.18) significantly so.
The two stars together therefore reinforce 
the growing trend for giant planets being more frequent around 
more metal-rich M dwarfs, and the 5~Jupiter mass Gl~676Ab being
found around a metal-rich star is consistent with the expectation
that the most massive planets preferentially form in disks with 
large condensate masses. 
}

\date{}


\titlerunning{Two massive planets around M0 dwarfs}
\authorrunning{Forveille et al.}

\maketitle

\section{Introduction}
Much recent theoretical work has gone into examining how planet
formation depends on stellar mass, because stellar 
mass significantly changes the physical conditions which
control the formation of planets.
A comparison, for instance, of the planet populations around Sun-like
stars on one hand, and around M dwarfs on the other hand, probes the
sensitivity of the planetary formation process to several physical
parameters: around lower mass stars gravity (hence disk rotation speed),
temperature (which regulates the position of the ice line) are both
lower, and, perhaps most importantly, disk mass scales approximately 
linearly with stellar mass \citep[e.g.][]{Scholz2006}. 

Within the ``core accretion'' paradigm, \citet[][]{Laughlin2004},
\citet[][]{Ida2005}, and \citet[][]{Kennedy2008} all predict 
that giant planet formation is inhibited around very-low-mass 
stars, while Neptune-mass planets should inversely be common. 
Within the same paradigm, but assuming that the properties of 
protoplanetary disks, contrary to observations, do not change
with stellar mass, \citet[][]{Kornet2006} predict instead that 
Jupiter-mass planets become more frequent in inverse
proportion to the stellar mass. Finally,  \citet[][]{Boss2006}
examines how planet formation depends
on stellar mass for planets formed by disk instability,
and concludes that the frequency of Jupiter-mass planet is
largely independent of stellar mass, as long as disks are massive
enough to become unstable.
One needs to note, though, that proto-planetary disks of a realistic
mass are likely to be gravitationally stable out to at least
10~AU. Planets can thus form through gravitational instability only 
beyond that distance, in a separation range only skimmed by
radial velocity monitoring and probed mostly by microlensing 
searches and direct imaging. Massive planets formed by gravitational
instability and found well within 5~AU must thus then have migrated 
inward. How giant planets migrating in the massive disks needed for 
gravitational instability can escape accreting enough mass to become 
a brown dwarf ($>13~\mathrm{M_{Jup}}$) is unclear 
\citet[e.g.][]{Stamatellos2009,Kratter2010}

Observationally, just a dozen 
of the close to 400 planetary systems currently known from radial 
velocity monitoring, 
are centered around M dwarfs (M$<$0.6\Msol)
\footnote{http://exoplanet.eu/catalog-RV.php}.
This no doubt reflects in part a selection bias, since 
many more of the intrisically brighter solar-type stars than 
of the fainter M dwarfs have been searched for planets, but there 
is increasing statistical evidence 
\citep[e.g.][]{Bonfils2006,Endl2006,Johnson2007,Johnson2010} 
that M dwarfs also genuinely have fewer massive planets 
($\sim$\Mjup) than the more massive solar-type stars.
They may, on the other hand, and though no rigorous statistical
analysis has yet been performed for that planet population, have 
a larger prevalence of the harder to detect Neptune-mass and 
super-Earth planets: a quarter of the $\sim$30~planets with 
M~sin(i)~$<$~0.1\Mjup 
known to date orbit an M dwarf, when solar-type stars 
outnumber M dwarfs by an order of magnitude in planet-search 
samples. Conversely, the highest mass planets known around
M dwarfs are the M~sin(i)~=~2\Mjup Gl~876b \citep{Delfosse1998,Marcy1998}
and HIP79431b \citep{Apps2010}, and at a larger orbital separation 
of $\sim$3~AUs the M~=${\approx}$3.5\Mjup 
OGLE-2005-BLG-071Lb microlensing planet \citep{Dong2009}, 
when over two dozen planets with 
masses over 10\Mjup are known around solar type stars. The statistical 
significance of that difference however remains modest, since the 
M dwarfs searched for planets only number in the few hundreds, when the 
apparent fraction of these very massive planets is under 1\% around solar 
type stars.

We present here the detection of two giant planets around M0 dwarfs,
a M~sin(i)~=~0.354\Mjup planet around HIP~12961, and a 
M~sin(i)~=~4.87\Mjup planet around Gl~676A.  

\section{Stellar characteristics}

\begin{table}
\centering
\begin{tabular}{l@{}lccl}
\hline
 \multicolumn{2}{l}{\bf Parameter}
& \multicolumn{1}{c}{\bf Gl~676A} 
& \multicolumn{1}{c}{\bf HIP~12961} 
& {\bf Notes}
\\
\hline
Spectral Type   &                
                & M0V  
                & M0V 
                & (a) 
\\ 
V               &        
                & $9.585 \pm 0.006$   
                & $10.31 \pm 0.04$    
                & (b)
\\ 
J               &           
                & $6.711 \pm 0.020$
                & $7.558 \pm 0.021$
                & (c) 
\\
K$_s$           &           
                & $5.825 \pm 0.029$ 
                & $6.736 \pm 0.018$ 
                & (c)
\\
BC$_{K_s}$       & 
                & 2.73
                & 2.61
                & (d)
                \\
$\pi$           &[mas]          
                & $60.79 \pm  1.62$ 
                &  $43.45 \pm 1.72$ 
                & (e)
\\
Distance                &[pc]           
                & $16.45 \pm 0.44$ 
                & $23.01 \pm 0.91$\\
$M_V$           &               
                & $8.50 \pm 0.06$ 
                & $8.50 \pm 0.09$ \\
$M_{K_s}$        &               
                & $4.74 \pm 0.06$ 
                & $4.93 \pm 0.09 $\\
$M_{bol}$       &               
                & 7.47 
                & 7.54 \\
$L_\star$       & [$\mathrm{L_\odot}$]          
                &  0.082
                &  0.076 \\
%
$v\sin i$       & [km\,s$^{-1}$] 
                & 1.6  $\pm$ 1.0
                & 1.5  $\pm$ 1.0 \\
                
$[Fe/H]$        &               
                & $0.18 $ 
                & $-0.07 $
                & (f) \\
$M_\star$       & [$\Msol$]             
                & 0.71 
                & 0.67 
                & (g) \\
\hline
\end{tabular}
\caption{
\label{table:stellar}
Observed and inferred stellar parameters for 
Gl~676A and HIP~12961. \hspace{5mm} 
Notes: (a) \citet{Hawley1996} for Gl~676A, and \citet{NLTT} for
        HIP~12961;
       (b) \citet{Koen2002} for Gl~676A, and computed from the 
       TYCHO (B$_T$,V$_T$) photometry for HIP~12961;
       (c) \citet{2MASS2006};       
       (d) computed from $J-K_s$ with the \citet{Leggett2001}
           BC$_{K_s}$ vs $J-K_s$ relation;
       (e) \citet{VanLeeuwen2007};
       (f) computed from M$_{K_s}$ and $V-K_s$ using the
           \citet{Schlaufman2010} calibration;
       (g)  computed from M$_{K_s}$ using the \citet{Delfosse2000}
           calibration; both masses are at the upper edge of the
           validity range of that calibration, and they might therefore
           have somewhat larger errors than its 10\% dispersion.
}
\end{table}
Table~\ref{table:stellar} summarizes the properties of the two host
stars, which we briefly discuss below.

\subsection{HIP~12961}
HIP~12961 (also CD-23$\deg$1056, LTT~1349, NLTT~8966, SAO~168043) 
was not identified as a member of the 25~pc volume until the publication 
of the \cite{Hipparcos1997} catalog, and has attracted very little attention:
it is mentioned in just 4 literature references, and always as part 
of a large catalog. Because HIP~1291 does not figure in the 
\citet{Gliese1991} catalog, it was omitted from the \citet{Hawley1996} 
spectral atlas of the late-type nearby stars. SIMBAD shows an M0 
spectral type, which seems to trace back to a classification of untraceable
pedigree listed in the NLTT catalog \citep{NLTT}, while \citet{Stephenson1986} 
estimated a K5 type from low-dispersion objective prism photographic plates. 
The absolute magnitude and color of HIP~12961, M$_V$=8.50 and V$-$K=3.57, 
suggest that its older NLTT spectral type is closer to truth 
\citep[e.g.][]{Leggett1992}. We adopt this spectral type for the 
reminder of the paper, but note that a modern classification from 
a digital low resolution spectrum is desirable. HIP~12961 has fairly
strong chromospheric activity, with 90\% of stars with spectral types 
K7 to M1 in the HARPS radial velocity sample (which however reject
the most active stars) having weaker Ca$_{\mathrm {II}}$ H and K lines,
and just 10\% stronger lines.  
The 2MASS photometry
(Table~\ref{table:stellar}) and the \citet{Leggett2001} $J-K$
colour vs bolometric relation result in a K-band bolometric correction 
of $BC_K=2.61$, and together with the parallax in a 0.076 \Lsol luminosity.

\subsection{Gl~676A}
The Gl~676 system (also CCDM~J17302-5138) has been recognized as a 
member of the immediate solar neighborhood for much longer, figuring in the 
original \citet{Gliese1969} catalog of the 20~pc volume. It consequently 
has 15 references listed in SIMBAD, though none of those dedicates more than 
a few sentences to Gl~676. The system comprises Gl~676A (also CD-51~10924, 
HIP~85647, CPD-51~10396) and Gl~676B, with respective spectral types of
M0V and M3V \citep{Hawley1996} and separated by $\sim$50" on the sky. 
At the distance of the system this angular distance translates into 
an $\sim$800~AU projected separation, which is probably far enough 
that Gl~676B didn't strongly influence the formation of the planetary 
system of Gl~676A. Gl~676A is a moderately active star, with a
Ca$_{\mathrm {II}}$ H and K emission strength at the third quartile of
the cumulative distribution for stars with spectral types between 
K7 and M1 in the HARPS radial velocity sample.
The 2MASS photometry (Table~\ref{table:stellar}) and
the \citet{Leggett2001} $J-K$ colour-bolometric relation result
in a K-band bolometric correction of $BC_K=2.73$, and together with 
the parallax in a 0.082 \Lsol luminosity. The \citet{Delfosse2000}
K-band Mass-Luminosity relation results in masses of respectively
0.71 and 0.29\Msol for Gl~676A and Gl~676B. The former is at the edge
of the validity range of the \citet{Delfosse2000} calibration, and
might therefore have somewhat larger uncertainties than the 
$\sim$10\% dispersion in that calibration.

\section{HARPS Doppler measurements and orbital analysis}
We obtained measurements of Gl~676A and HIP~12961 with HARPS 
\citep[High Accuracy Radial velocity Planet Searcher][]{Mayor2003}, 
as part of the guaranteed-time program of the instrument consortium. 
HARPS is a high-resolution (R~=~115 000) fiber-fed echelle 
spectrograph, optimized for planet search programs and 
asteroseismology. It is the most precise spectro-velocimeter to 
date, with a long-term instrumental RV accuracy well under 
1~m\,s$^{-1}$ \citep[e.g.][]{Lovis2006,Mayor2009}.
When it aims for ultimate radial velocity precision, HARPS uses
simultaneous exposures of a thorium lamp through a calibration fiber.
For the present observations however, we relied instead on its excelent
instrumental stability (nightly instrumental drifts $<$~1\,m~s$^{-1}$).
Both targets are too faint for us to reach the stability limit
of HARPS within realistic integration times, and dispensing with the
simultaneous thorium light produces cleaner stellar spectra, more
easily amenable to quantitative spectroscopic analysis.
The two stars were observed as part of the volume-limited HARPS 
search for planets \citep[e.g.][]{Moutou2009,LoCurto2010}. 
While generally refered to as F-G-K stars, for the sake of 
concision, the targets of that program actually include M0 
dwarfs \citep[][]{LoCurto2010}.  

We used 15~mn exposures for both stars, obtaining median S/N ratios
(per pixel at 550~nm) of 53 for the V=9.58 Gl~676A, and 49 for the
V=10.31 HIP~12961. The 69 and 46 radial velocities of Gl~676A and 
HIP~12961 (Tables~\ref{TableRV_Gl676} and \ref{TableRV_HIP12961}, 
only available electronically) were obtained with the standard HARPS 
reduction pipeline, based on cross-correlation 
with a stellar mask and on a precise nightly
wavelength calibration from ThAr spectra \citep{Lovis2007}.
The median internal errors of these velocities are respectively 1.9 and 
2.8~m\,s$^{-1}$, and include a $\sim$~0.2~m\,s$^{-1}$ noise on the 
nightly zero-point measurement, a $\sim$~0.3~m\,s$^{-1}$ uncertainty 
on the instrumental  drift, and the photon noise computed from 
the full Doppler information content of the spectra 
\citep{Bouchy2001}. The photon noise contribution completely 
dominates the error budget for these moderately faint sources.

\subsection{A Saturn-mass planet around HIP~12961}
\begin{figure}
\centering
\includegraphics[width=0.9\linewidth]{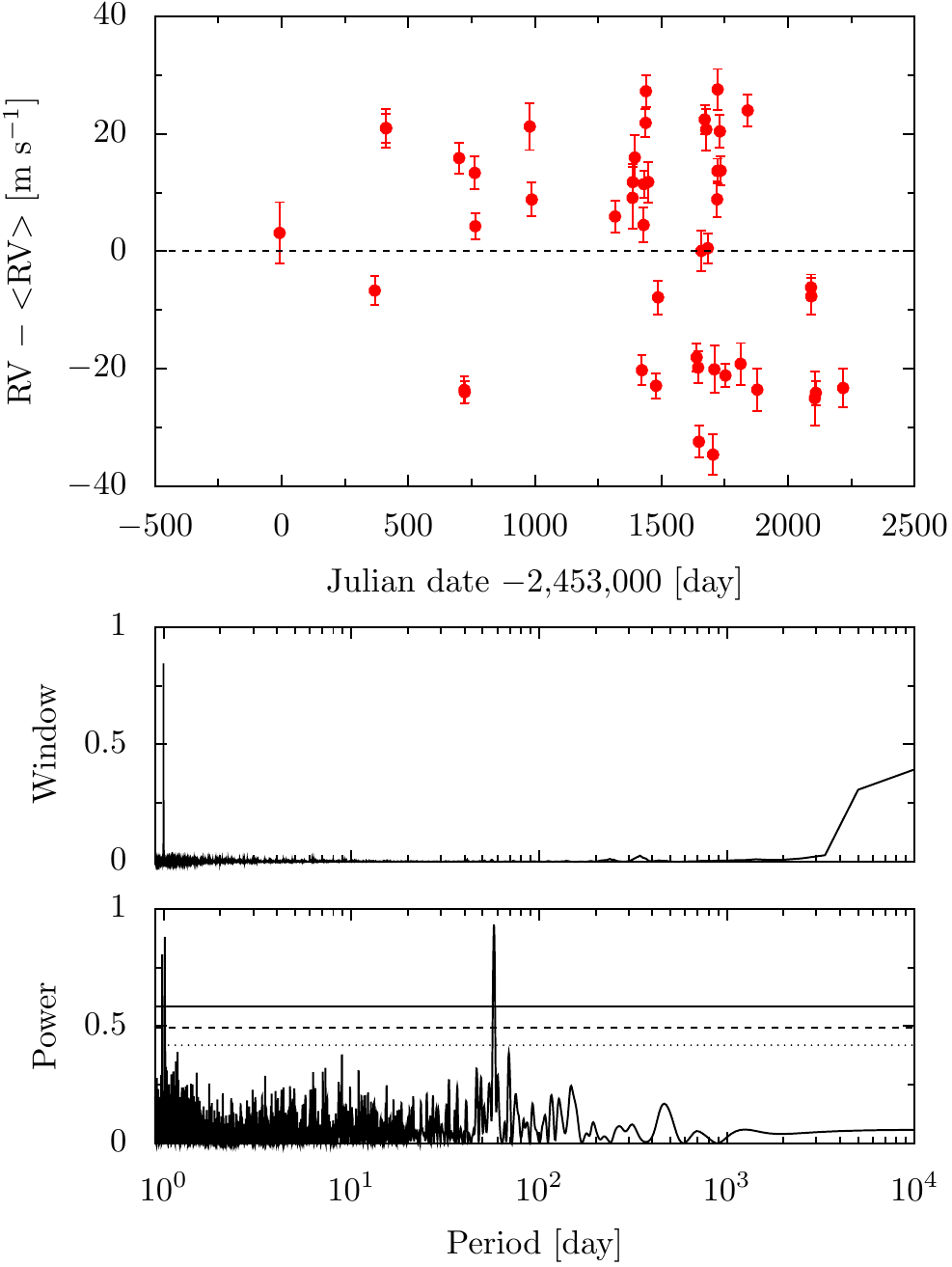}
       \caption{
        HARPS radial velocities of HIP~12961 as a function of barycentric 
        Julian day (top panel), spectral power window (middle panel)
        and Lomb-Scargle periodogram of these
        velocities (bottom panel). The horizontal lines mark false 
        alarm probabilities equivalent to  1, 2 and 3 sigmas 
        significance levels for Gaussian noise.
        }
       \label{Fig_HIP12961}
\end{figure}
\begin{figure}
\centering
\includegraphics[width=0.9\linewidth]{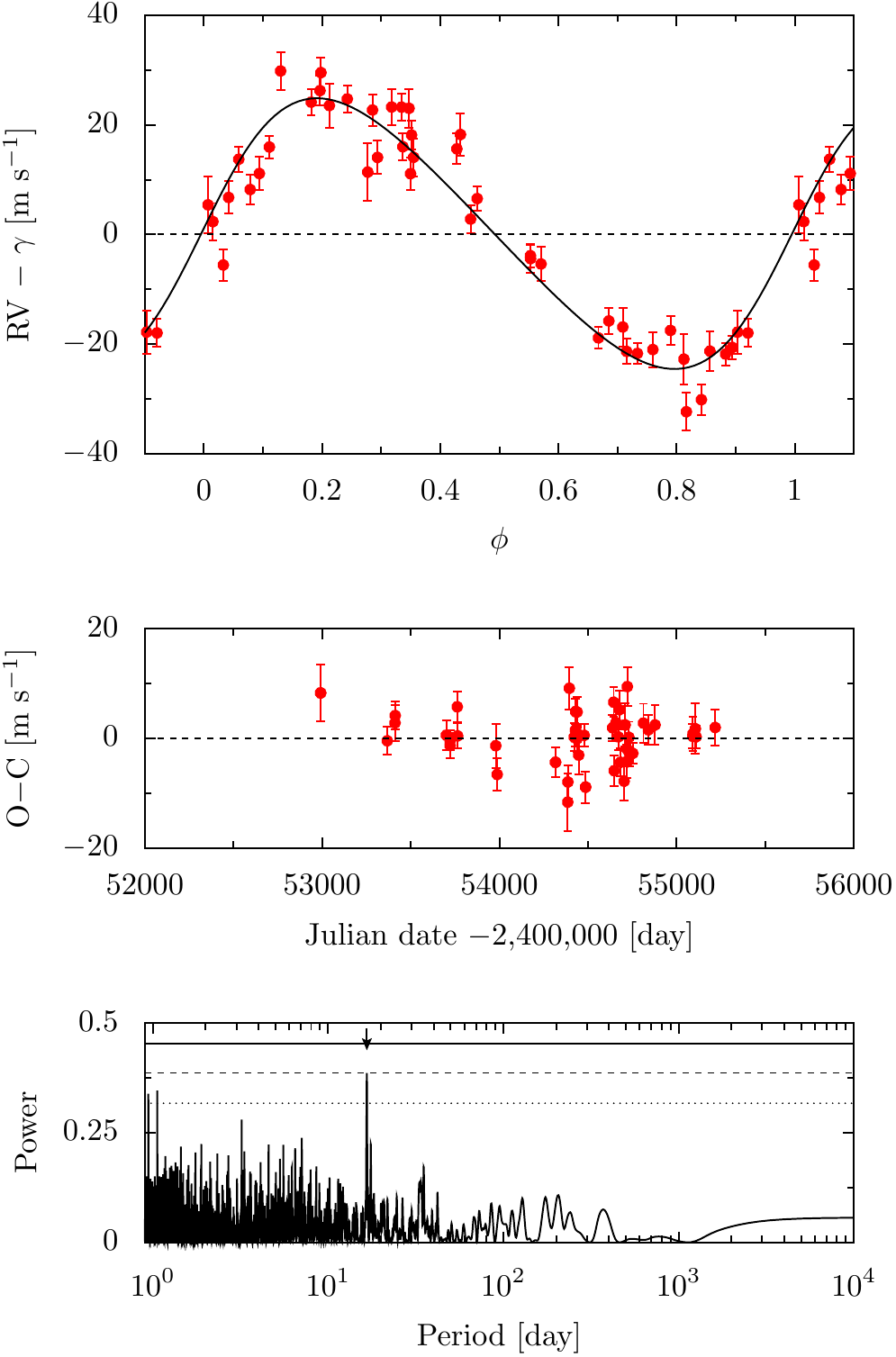}
       \caption{
        HARPS radial velocities of HIP~12961 phased to the 57.44~days
        period, overlaid with the adjusted Keplerian orbit
        (top panel). Residuals from the Keplerian orbit 
        as a function of barycentric Julian day (middle panel), 
        and Lomb-Scargle periodogram of these
        residuals (bottom panel). The horizontal lines mark false 
        alarm probabilities equivalent to  1, 2 and 3 sigmas 
        significance levels for Gaussian noise.
       }
       \label{Fig_HIP12961-orbit}
\end{figure}

The computed radial velocities of HIP~12961 vary with a 
$\sim$~60$m\,s^{-1}$ peak to peak 
amplitude (Fig.~\ref{Fig_HIP12961}, top panel),
an order of magnitude above their 2.6~$m\,s^{-1}$ average 
photon noise, and well above the $\lesssim$10 $m\,s^{-1}$
maximum jitter expected from the chromospheric activity. 
The variations show no correlation with the 
bisector span (rms 6$m\,s^{-1}$), the depth (10.65\%, 
with 0.07\% rms) or width (3.566~km$\,s^{-1}$, with 14$m\,s^{-1}$
rms) of the correlation profile, or any of the standard 
stellar activity diagnostic, making orbital motion by far 
their most likely cause. The Lomb-Scargle periodogram of 
the velocities shows one highly significant peak at 57.45~days 
(Fig.~\ref{Fig_HIP12961}, middle and
bottom panel), as well as its 4 aliases at +-1 sidereal and 
civil days. Phasing of the velocities on that period 
shows well sampled smooth variations (Fig.~\ref{Fig_HIP12961-orbit} 
top panel). A Keplerian fit (Table~\ref{TableElements})
yields a moderately eccentric orbit ($e$~=~0.2) with a
25~m\,s$^{-1}$ semi-amplitude.

The rms amplitude of the 
residuals from that orbit is 3.8~m\,s$^{-1}$
(Fig.~\ref{Fig_HIP12961-orbit},
middle panel), significantly
above the 2.6~m\,s$^{-1}$ average photon noise. The square
root ot the reduced ${\chi}^2$ of the fit is consequently 1.5.
The radial velocities may therefore contain information beyond
the detected planet, but the highest peak in the periodogram 
of the radial velocity residuals (16.6~days) only has
2~$\sigma$ significance. It also coincides with a signal in
the periodogram of the correlation profile's depth, and it is broadly
consistent with the stellar rotation period expected from the significant
chromospheric activity. This peak, if not just noise, is therefore much 
more likely to reflect rotational modulation of stellar spots than 
a planet. There is no current evidence for additional planets in the system.

Together with the 0.67~$\Msol$ (Table~\ref{table:stellar}) stellar mass, 
the orbital parameters translate into a minimum companion mass of
0.35$\mathrm{M_{Jup}}$, or 1.2 Saturn-mass, with a 0.25~AU
semi-major axis. For an Earth-like albedo of 0.35 the equilibrium 
temperature at that distance from a 0.076 \Lsol (Table~\ref{table:stellar})
luminosity star is 263~K, slightly higher than the terrestrial 255~K
but below the $\sim$270~K threshold for triggering a runaway greenhouse 
effect \citet{Selsis2007}. A putative massive moon of HIP~12961b 
could therefore be potentially hospitable to life.

The {\it a priori}
geometric probability that HIP~12961b  transits across HIP~12961
is approximately {1\%}. As usual for planets of M dwarfs, the transit 
would be deep ($\sim$2.5\%) and therefore well suited to high quality
transmission spectroscopy of the planetary atmosphere, as well as to
searches for transits by planetary moons. This high potential return 
offsets the long odds to some extent, and the deep transits
would be within easy reach of amateur-grade equipment. The star will 
thus be well worth searching for transits, once 
additional radial velocity measurements will have narrowed
down the time windows for potential planetary transits. 

\subsection{A massive long period planet around Gl~676A}
\begin{figure}
\includegraphics[width=9cm,angle=0]{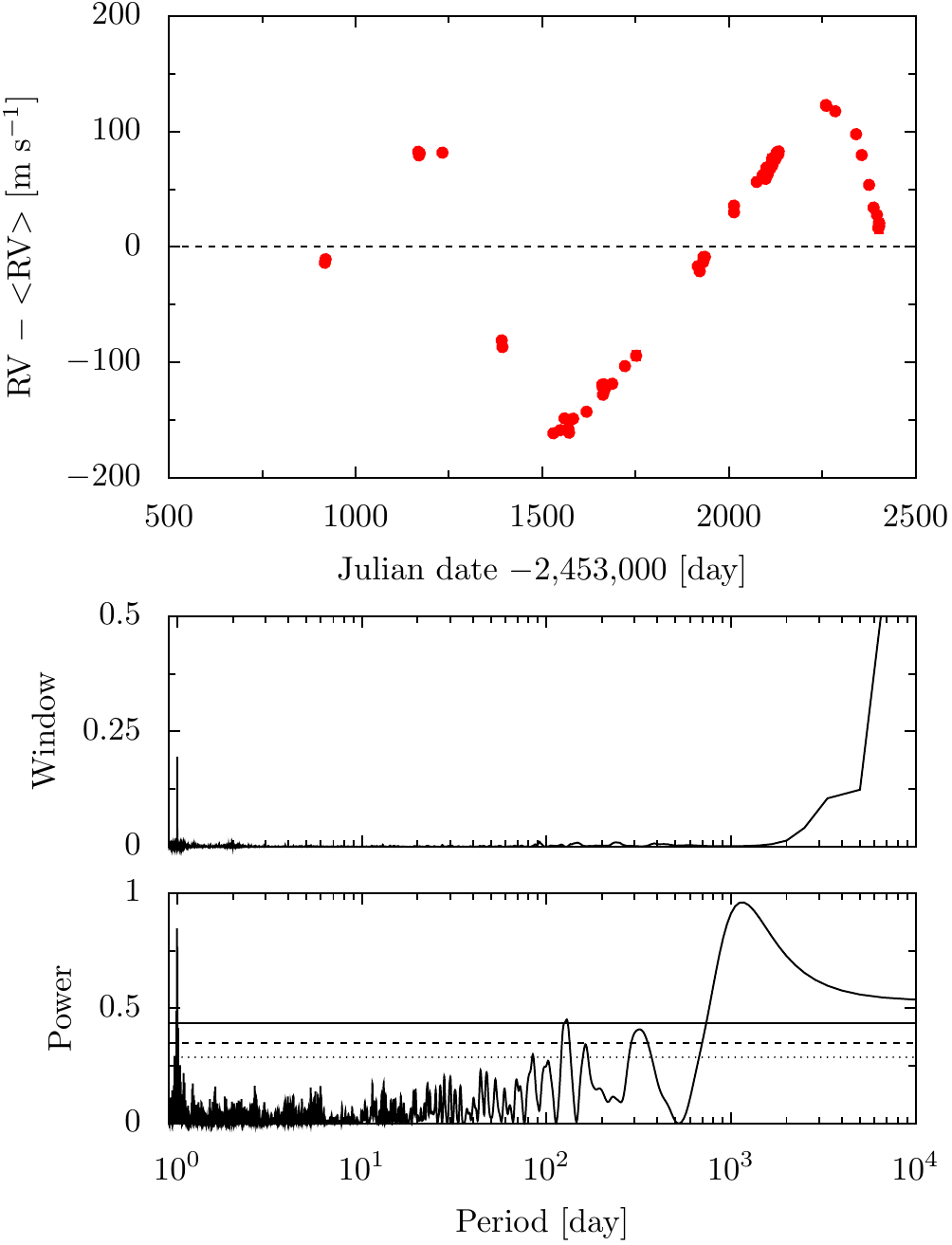}
\caption{
        HARPS radial velocities of Gl~676A as a function of barycentric 
        Julian day (top panel), spectral power window (middle panel)
        and Lomb-Scargle periodogram of these
        velocities (bottom panel). The horizontal lines mark false 
        alarm probabilities equivalent to  1, 2 and 3 sigmas 
        significance levels for Gaussian noise.
}
\label{Fig_Gl676A}
\end{figure}
The computed velocities of Gl~676A (Table~\ref{TableRV_Gl676}) exhibit 
unambiguous variations of several hundred m\,s$^{-1}$ with a period 
slightly over our current observing span, superimposed upon a slower 
drift (Fig.\ref{Fig_Gl676A}, upper panel). 
The correlation profile depth (13.61\%, with 0.12\% rms), its width,
the bisector span (5~m\,s$^{-1}$ rms), and the chromospheric indices 
show no systematic variations that would correlate with the radial
velocity changes. 

A fit of a Keplerian plus a constant acceleration to the radial 
velocities (Table~\ref{TableElements}) yields a period of 
1056.8~$\pm$~2.8~days, a semi-amplitude of 
122.8~$\pm$~1.9~m\,s$^{-1}$, and a 10.7~m\,s$^{-1}$yr$^{-1}$
acceleration. 
\begin{figure}
\includegraphics[width=9cm,angle=0]{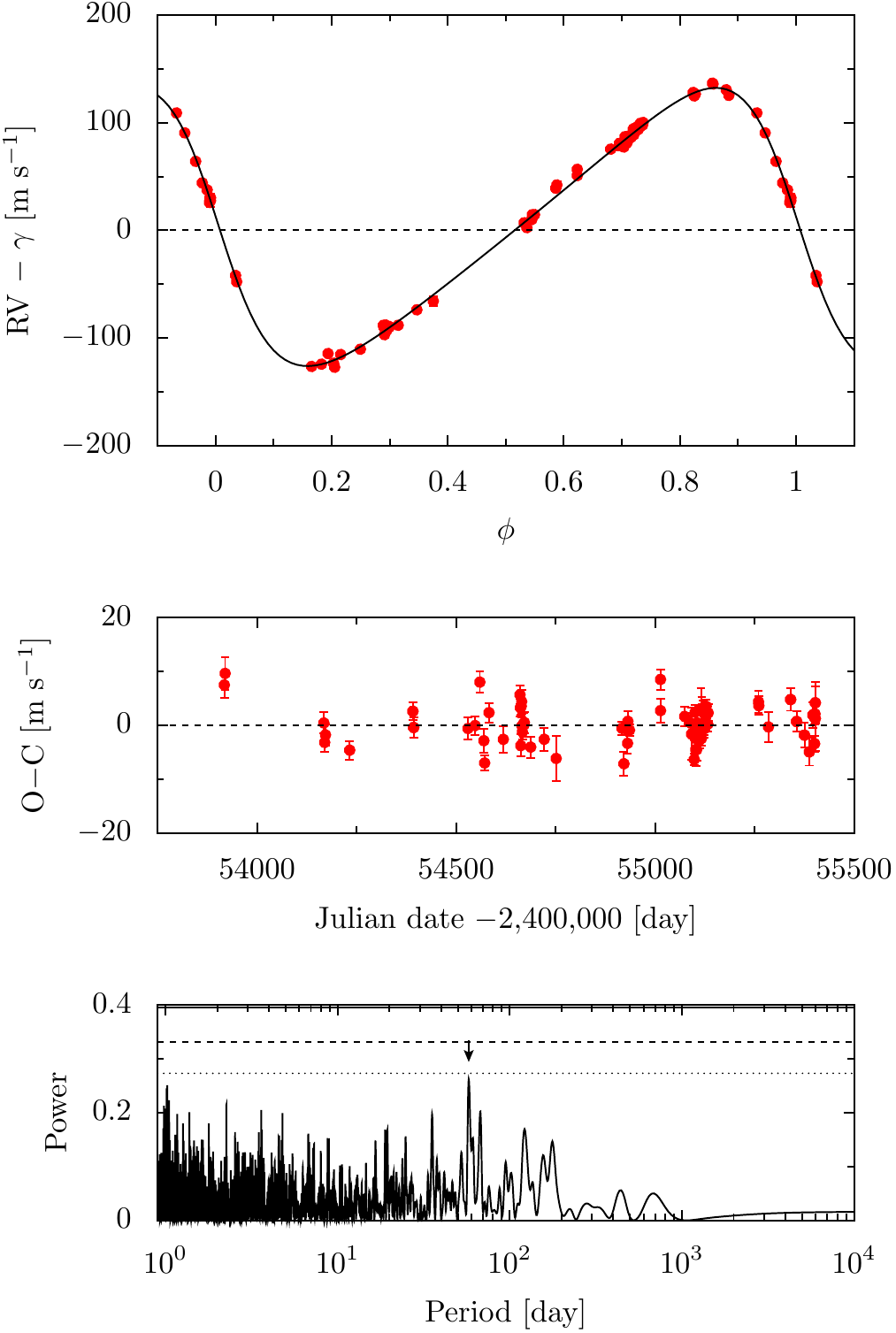}
\caption{
Top panel: HARPS radial velocities of 
Gl~676A as a function of orbital phase, after subtraction 
of adjusted linear drift and overlaid with the 
adjusted Keplerian orbit. Middle panel: residual of the HARPS radial 
from the adjusted orbit and linear drift, as a function of time.
Bottom Panel: Lomb-Scargle periodogram of the residuals
The horizontal lines mark false alarm probabilities equivalent 
to  1 and 2 sigmas significance levels for Gaussian noise.
}
\label{Fig_Gl676A-orbit}
\end{figure}

Together with the 0.71~$\Msol$ (Table~\ref{table:stellar}) 
stellar mass, the orbital parameters imply a companion mass of 
4.9 $\mathrm{M_{Jup}}$ 
and a 1.82~AU semi-major axis. At the 16.5~pc
distance of the Gl~676 system, the minimum astrometric
wobble of Gl~676A, for a sin(i)~=~1 edge-on orbit,
is $\pm$0.67~mas. This is within reach of both the FGS
instrument on {\it HST} \citep[e.g.][]{Martioli2010} and
imagers on 8m-class telescopes \citep{Lazorenko2009}.
The companion of Gl~676A therefore belongs to
the small group of non-eclipsing planets for which
the inclination ambiguity can potentially be lifted 
with existing instruments. We have started such
astrometric observations using the {\it FORS2} imager 
of the ESO {\it VLT}.

The line of sight acceleration of Gl~676A by its 0.3\Msol
Gl~676B stellar companion is of order $G\,M_B/r_{AB}^2$, and 
from the 800~AU projected separation is therefore under 
0.1~m\,s$^{-1}$yr$^{-1}$.   
The two orders of magnitude discrepancy between the observed 
acceleration and that expected from Gl~676B demonstrates that 
the system contains an additional massive body. That body
could be planetary if its separation is under $\sim$15~AU
(0.9''), stellar if that separation is above $\sim$40~AU
(2.4''), or a brown dwarf for intermediate separations.
Adaptive optics imaging could easily narrow down these
possibilities, as will the continuing radial velocity 
monitoring to constrain the curvature of the radial
velocity trend, and the on-going astrometric effort.

The rms amplitude of the residuals around the 
Keplerian+acceleration orbit
is 3.4~m\,s$^{-1}$, significantly above our 1.7~~m\,s$^{-1}$
average measurement error. The square root of the reduced 
${\chi}^2$ of the fit is consequently 2.0, indicating that the 
residuals contain structure above the photon noise. The highest peak 
in a Lomb-Scargle \citep{Lomb1976,Scargle1982} periodogram 
of the residuals however only rises to a level equivalent 
to a 1~$\sigma$ detection (Fig.\ref{Fig_Gl676A-orbit}). There is 
therefore no immediate evidence for additional planets in the
system. The excess residuals may simply reflect jitter
from the moderate stellar activity of Gl~676A, or alternatively
they could be early signs of multiple additional planets, which additional
observations would then eventually disentangle.

\begin{table}
\caption{Orbital elements for the Keplerian
orbital models of HIP~12961 and Gl~676A.}
\label{TableElements}
\centering
\begin{tabular}{c c c}
\hline\hline
{\bf Element} & {\bf HIP~12961} & {\bf Gl~676A} \\
\hline
$\gamma$ [km\,s$^{-1}$]&  33.0463  $\pm$  0.0014 & -39.108 $\pm$ 0.091 \\
$d{\gamma}/dt$ [m\,s$^{-1}$\,yr$^{-1}$]&  --  &  10.66 $\pm$  0.61 \\
Epoch [BJD]&  --  &  2450000 \\
P [days] &  57.435    $\pm$ 0.042 & 1056.8 $\pm$   2.8 \\
e  &         0.166    $\pm$ 0.034 & 0.326 $\pm$   0.009 \\
$\omega$  [deg.] & 272 $\pm$  13 & 85.7 $\pm$   1.4 \\
T0  [BJD] &    2454428.4 $\pm$   2.0 & 2455411.9 $\pm$  3.0 \\
K1  [m\,s$^{-1}$] &    24.71 $\pm$     0.86  &  129.3 $\pm$     1.2 \\
\hline
a          [AU]   &    0.25   &     1.82  \\
M~sin(i)  [\Mjup] &    0.35   &  4.9      \\
\hline
$\sigma$(O-C) [m\,s$^{-1}$] &  3.9    &  3.4 \\
N$_{mes}$       &   46   &  69 \\
N$_{par}$       &   6   &   7 \\
$\chi^{2}_{r}$  &   2.3   &  3.9 \\
\hline

\end{tabular}
\end{table}

\section{Discussion}
As discussed above, HIP~12961 and Gl~676A are orbited by giant
planets with minimum masses of approximately 0.5 and 5 Jupiter 
masses. The latter is twice the M~sin(i)~=~2\Mjup of Gl~876b 
\citep{Delfosse1998,Marcy1998} and HIP79431b \citep{Apps2010},
previously the highest mass planets found by radial velocity
monitoring of M~dwarfs, and above the 3.8 or 3.4~\Mjup
(from two degenerate solutions) of the OGLE-2005-BLG-071Lb 
\citep{Dong2009} microlensing planet. The M0V Gl~676A however
is significantly more massive (0.71~\Msol, Table~\ref{table:stellar}) 
than the M4V Gl~876 \citep[0.33~\Msol][]{Correia2010}
and the M3V HIP79431 \citep[0.49~\Msol][]{Apps2010}. The higher
mass of its planet therefore remains in approximate line with
the current upper envelope of the planetary versus stellar mass 
diagram. These most massive planets are rare at any stellar
mass, with an occurence rate under 1\%, suggesting that they
can form only under the most favorable conditions. They have been
suggested to form through gravitational instability, with
their lower mass counterparts forming by core accretion.
Proto-planetary disks of any realistic mass, however, are expected 
be gravitationally stable out to beyond 10~AU. If Gl~676Ab 
formed through gravitational instability, it would therefore 
have undergone much inward migration, through a very massive 
disk. How it could escape accreting enough mass during this
migration to become a brown dwarf is unclear.

Gl~676A and HIP~12961 increase the sample of M dwarfs with giant 
planets (Saturn-mass and above) from 7 to 9, and therefore
offer an opportunity to evaluate the trend
\citep{Johnson2009,Schlaufman2010} for giant planets being more 
common around more metal-rich M~dwarfs.
Adopting the very recent \cite{Schlaufman2010} metallicity calibration  
of the M$_{K_s}$ vs V-K$_S$ plane, which finds metallicities
approximately half-way between those of the earlier
\citet{Bonfils2005} and \citet{Johnson2009} calibrations, 
the metallicities of Gl~676A and HIP~12961 are $0.18$ 
and $-0.07$. Both values are above the [Fe/H]~=~-0.17 average 
metallicity for the solar neighborhood in the \cite{Schlaufman2010} 
metallicity scale, the latter very significantly so. The
two new planets therefore clearly reinforce the incipient trend,
and help suggest that more massive planets are found around more
metal-rich M-dwarfs.

\begin{acknowledgements}

We would like to thank the ESO La Silla staff for their excellent support,
and our collaborators of the HARPS consortium for making this instrument 
such a success, as well as for contributing some of the observations
through observing time exchange. 
TF thanks the Institute for Astronomy of the University of Hawaii for
its kind hospitality while much of this paper was written.
Financial support from the "Programme National de 
Plan\'etologie'' (PNP) of CNRS/INSU, France, is gratefully acknowledged.

XB acknowledge support from the Funda\c{c}\~ao para a Ci\^encia
e a Tecnologia (Portugal) in the form of a fellowship (reference
SFRH/BPD/21710/2005) and a program (reference
PTDC/CTE-AST/72685/2006), as well as the Gulbenkian Foundation for
funding through the ``Programa de Estímulo à Investigação''.

NCS acknowledges the support by the European Research Council/European 
Community under the FP7 through a Starting Grant, as well as in the 
form of grants reference PTDC/CTE-AST/66643/2006 and 
PTDC/CTE-AST/098528/2008, funded by Funda\c{c}\~ao para a Ci\^encia e 
a Tecnologia (FCT), Portugal. NCS would further like to thank the 
support from FCT through a Ci\^encia\,2007 contract funded by 
FCT/MCTES (Portugal) and POPH/FSE (EC).

\end{acknowledgements}


\bibliographystyle{aa}
\bibliography{biblio}

\begin{table}
\caption{Radial-velocity measurements and error bars for Gl~676A
All values are relative to the solar system barycenter, and
corrected from the small perspective acceleration using the Hipparcos
parallax and proper motion. Only available electronically.}
\label{TableRV_Gl676}
\centering
\begin{tabular}{c c c}
\hline\hline
\bf JD-2400000 & \bf RV & \bf Uncertainty \\
 & \bf [km\,s$^{-1}$] & \bf [km\,s$^{-1}$] \\
\hline
53917.747997  &  -39.097817  &  0.002411  \\
53919.735174  &  -39.094780  &  0.003022  \\
54167.897856  &  -39.001569  &  0.001963  \\
54169.895854  &  -39.004708  &  0.001784  \\
54171.904445  &  -39.002836  &  0.001928  \\
54232.818013  &  -39.002344  &  0.001722  \\
54391.491808  &  -39.165064  &  0.001699  \\
54393.489934  &  -39.170823  &  0.001861  \\
54529.900847  &  -39.245590  &  0.002076  \\
54547.915016  &  -39.242932  &  0.001751  \\
54559.815698  &  -39.232718  &  0.002002  \\
54569.903637  &  -39.241409  &  0.002270  \\
54571.889460  &  -39.245083  &  0.001360  \\
54582.820292  &  -39.232950  &  0.001791  \\
54618.755585  &  -39.226941  &  0.002436  \\
54660.661636  &  -39.203487  &  0.001727  \\
54661.772229  &  -39.205488  &  0.001626  \\
54662.675237  &  -39.212174  &  0.002052  \\
54663.811590  &  -39.204402  &  0.001537  \\
54664.790043  &  -39.203176  &  0.002240  \\
54665.786377  &  -39.207272  &  0.001557  \\
54666.696058  &  -39.208065  &  0.001421  \\
54670.672602  &  -39.205235  &  0.002207  \\
54671.603329  &  -39.204414  &  0.001945  \\
54687.561959  &  -39.202682  &  0.001968  \\
54721.554874  &  -39.187358  &  0.002167  \\
54751.490690  &  -39.178373  &  0.004248  \\
54916.819805  &  -39.100792  &  0.001226  \\
54921.892971  &  -39.105131  &  0.002259  \\
54930.906849  &  -39.097319  &  0.001873  \\
54931.795103  &  -39.092830  &  0.001922  \\
54935.817789  &  -39.092687  &  0.001131  \\
55013.686615  &  -39.048271  &  0.001909  \\
55013.743720  &  -39.054037  &  0.002253  \\
55074.520060  &  -39.027729  &  0.001871  \\
55090.507026  &  -39.021867  &  0.001839  \\
55091.528800  &  -39.023418  &  0.004842  \\
55098.494144  &  -39.025046  &  0.001144  \\
55100.540947  &  -39.015400  &  0.001408  \\
55101.490472  &  -39.019816  &  0.002182  \\
55102.502862  &  -39.021444  &  0.003021  \\
55104.540258  &  -39.020592  &  0.002090  \\
55105.523635  &  -39.016518  &  0.003934  \\
55106.519974  &  -39.014727  &  0.001956  \\
55111.509339  &  -39.015782  &  0.001428  \\
55113.497880  &  -39.013708  &  0.001459  \\
55115.514997  &  -39.008140  &  0.003762  \\
55116.487535  &  -39.013302  &  0.001343  \\
55117.493046  &  -39.007306  &  0.002138  \\
55121.526645  &  -39.006189  &  0.002180  \\
55122.505321  &  -39.008620  &  0.001979  \\
55124.497834  &  -39.008016  &  0.001203  \\
55127.516794  &  -39.005698  &  0.001163  \\
55128.513957  &  -39.002138  &  0.001187  \\
55129.495404  &  -39.002679  &  0.001308  \\
55132.495755  &  -39.003770  &  0.001430  \\
55133.493189  &  -39.001253  &  0.001564  \\
55259.907275  &  -38.961036  &  0.002051  \\
55260.864406  &  -38.961729  &  0.001725  \\
55284.893135  &  -38.966427  &  0.002793  \\
55340.708504  &  -38.986337  &  0.002081  \\
55355.795443  &  -39.004416  &  0.001919  \\
55375.610729  &  -39.030286  &  0.002274  \\
55387.656686  &  -39.049946  &  0.002556  \\
55396.537980  &  -39.056201  &  0.002343  \\
55400.642866  &  -39.067768  &  0.001486  \\
55401.594785  &  -39.063691  &  0.001907  \\
55402.590925  &  -39.066073  &  0.005969  \\
55402.702771  &  -39.063337  &  0.003840  \\
\hline
\end{tabular}

\end{table}

\begin{table}
\caption{Radial-velocity measurements and error bars for HIP~12961
All values are relative to the solar system barycenter, and
corrected from the small perspective acceleration using the Hipparcos
parallax and proper motion. Only available electronically.}
\label{TableRV_HIP12961}
\centering
\begin{tabular}{c c c}
\hline\hline
\bf JD-2400000 & \bf RV & \bf Uncertainty \\
 & \bf [km\,s$^{-1}$] & \bf [km\,s$^{-1}$] \\
\hline
52991.634308  &  33.05175  &  0.00524  \\
53367.633702  &  33.04193  &  0.00252  \\
53411.567881  &  33.06961  &  0.00326  \\
53412.538676  &  33.06959  &  0.00249  \\
53700.702601  &  33.06450  &  0.00265  \\
53721.606360  &  33.02500  &  0.00226  \\
53722.659901  &  33.02462  &  0.00182  \\
53762.549717  &  33.06197  &  0.00282  \\
53764.526469  &  33.05289  &  0.00228  \\
53979.914378  &  33.06988  &  0.00401  \\
53987.805086  &  33.05744  &  0.00290  \\
54316.859111  &  33.05455  &  0.00275  \\
54385.727404  &  33.05773  &  0.00529  \\
54386.677499  &  33.06041  &  0.00303  \\
54394.741029  &  33.06459  &  0.00384  \\
54422.684669  &  33.02836  &  0.00258  \\
54429.638114  &  33.05310  &  0.00294  \\
54430.626508  &  33.06007  &  0.00231  \\
54437.666007  &  33.07050  &  0.00235  \\
54438.616998  &  33.07589  &  0.00277  \\
54447.623874  &  33.06043  &  0.00345  \\
54478.620266  &  33.02571  &  0.00210  \\
54486.564941  &  33.04077  &  0.00287  \\
54638.924339  &  33.03056  &  0.00237  \\
54644.919198  &  33.02882  &  0.00270  \\
54647.921380  &  33.01618  &  0.00274  \\
54657.875027  &  33.04869  &  0.00339  \\
54670.938132  &  33.07108  &  0.00248  \\
54676.938113  &  33.06938  &  0.00353  \\
54682.924974  &  33.04915  &  0.00254  \\
54703.900626  &  33.01398  &  0.00347  \\
54708.869197  &  33.02851  &  0.00400  \\
54719.829342  &  33.05747  &  0.00304  \\
54720.805422  &  33.06231  &  0.00208  \\
54721.896983  &  33.07619  &  0.00349  \\
54730.830908  &  33.06906  &  0.00285  \\
54733.780908  &  33.06236  &  0.00242  \\
54752.787477  &  33.02746  &  0.00194  \\
54812.601219  &  33.02945  &  0.00354  \\
54840.603108  &  33.07261  &  0.00276  \\
54878.515553  &  33.02504  &  0.00364  \\
55090.820449  &  33.04247  &  0.00219  \\
55091.853089  &  33.04096  &  0.00308  \\
55105.740055  &  33.02358  &  0.00457  \\
55109.787500  &  33.02451  &  0.00205  \\
55217.599913  &  33.02533  &  0.00323  \\
\hline
\end{tabular}
\end{table}

\end{document}